\documentclass[aps,twocolumn,superscriptaddress,amssymb]{revtex4-1}

\usepackage{amsmath}

\usepackage{graphicx}
\usepackage{floatrow}
\usepackage[leftcaption]{sidecap}
\sidecaptionvpos{figure}{right}
\usepackage{amsmath,amsfonts,amsbsy,amssymb,array,accents,dsfont}
\usepackage{enumerate,array,latexsym,graphicx,mathrsfs,verbatim,psfrag}
\usepackage{bm} 
\usepackage[normalem]{ulem}
\usepackage{multirow}
\usepackage{cancel}
\usepackage{chngpage} 
\usepackage{nccmath}
\usepackage{booktabs} 
\usepackage[usenames]{color}
\usepackage[utf8]{inputenc}
\usepackage[english]{babel}
\usepackage{fancybox}
\usepackage[dvipsnames]{xcolor}

\usepackage{tikz}
\usepackage{pgfplots}
\usetikzlibrary{decorations.text,intersections, pgfplots.fillbetween}
\usetikzlibrary{math}
 \usetikzlibrary{snakes,3d,shapes.geometric,shadows.blur}
\usepackage{tikz-3dplot}

\usepackage{sidecap}

\sidecaptionvpos{figure}{c}

\usepackage[colorlinks=true,      linkcolor=blue,      urlcolor=blue,      
            filecolor=blue,      citecolor=blue,       pdfstartview=FitH,     
						pdfpagemode=UseNone,      bookmarksopen=true]{hyperref}  
\usepackage[all]{hypcap}     

\makeatletter
\renewcommand{\p@subsection}{}
\renewcommand{\p@subsubsection}{}
\makeatother

\newcommand{\captionfonts}{\small}
\makeatletter  
\long\def\@makecaption#1#2{%
  \vskip\abovecaptionskip
  \sbox\@tempboxa{{\captionfonts #1: #2}}%
 \ifdim \wd\@tempboxa >\hsize
    {\captionfonts #1: #2\par}
  \else
    \hbox to\hsize{\hfil\box\@tempboxa\hfil}%
  \fi
  \vskip\belowcaptionskip}
\makeatother   

\DeclareMathSymbol{\medhatsym}{\mathord}{largesymbols}{"62} 

\DeclareMathSymbol{\medtildesym}{\mathord}{largesymbols}{"65}

\newcommand{\comm}[1]{} 

\def\IR{\mathbb{R}}

\def\({\left(}
\def\){\right)}
\def\[{\left[}
\def\]{\right]}

\def\One{{\hbox{ 1\kern-.8mm l}}}

\def\barray{\begin{array}}
\def\earray{\end{array}}
\def\be{\begin{equation}}
\def\ee{\end{equation}}
\def\bea{\begin{eqnarray}}
\def\eea{\end{eqnarray}}
\def\bal{\begin{align}}
\def\eal{\end{align}}
\def\nn{\nonumber}

\def\-{\,-\,}
\def\={\,=\,}
\def\+{\,+\,}

\numberwithin{equation}{section} 

\makeatletter
\g@addto@macro\bfseries{\boldmath}
\makeatother

\definecolor{cardinal}{rgb}{0.6,0,0}
\definecolor{darkgreen}{rgb}{0,0.4,0}
\definecolor{purple}{rgb}{0.5, 0, 0.5}
\definecolor{golden}{rgb}{0.92, 0.7, 0}
\definecolor{midnight}{rgb}{0, 0, 0.5}
\definecolor{darkblue}{rgb}{0, 0, 0.8}

\def\IR{\mathds{R}}

\def\cV{{\cal V}}

\newcommand*{\Scale}[2][4]{\scalebox{#1}{$#2$}}%

\usetikzlibrary{positioning,calc,fadings,decorations.pathreplacing}

\tikzset{
 diffuse color/.initial = black,                       
}

\tikzset{
 linear opacity/.initial=0.5,                          
 linear stroke/.style = {                              
   preaction={                                         
     draw=\pgfkeysvalueof{/tikz/diffuse color},        
     line width = (2.0-#1)*\pgflinewidth,              
     opacity=\pgfkeysvalueof{/tikz/linear opacity},white}},  
 diffuse gradient/.style={                             
   draw = none,                                        
   linear opacity=#1,                                  
   linear stroke/.list={0.0,#1,...,1.0}},              
 diffuse gradient/.default=1,                          
}

\tikzset{
 non-linear stroke/.style = {                          
   preaction={                                         
     draw=\pgfkeysvalueof{/tikz/diffuse color},        
     line width = (2.0-#1)*\pgflinewidth,              
     opacity=#1,white}},                                     
 diffuse falloff/.style={                              
   draw = none,                                        
   non-linear stroke/.list={0.0,#1,...,1.0}},          
 diffuse falloff/.default=1,                           
}
 
\tikzset{%
  >=latex, 
  inner sep=0pt,%
  outer sep=2pt,%
  mark coordinate/.style={inner sep=0pt,outer sep=0pt,minimum size=3pt,
    fill=black,circle}%
}

\usepackage{titlesec}
\titlespacing*{\section}
{0pt}{0.5cm}{0.2cm}
\titlespacing*{\subsection}
{0pt}{0.25cm}{0.1cm}

\begin{document}

\title{Geometric Resolution of Schwarzschild Horizon}

\author{Ibrahima Bah}
\email{iboubah@jhu.edu}
\author{Pierre Heidmann}
\email{pheidma1@jhu.edu}
\affiliation{Department of Physics and Astronomy, Johns Hopkins University, 3400 North Charles Street, Baltimore, MD 21218, USA}

\begin{abstract}

We provide the first example of a geometric transition that resolves the Schwarzschild black hole into a smooth microstructure in eleven-dimensional supergravity on a seven-torus. The geometry is indistinguishable from a Schwarzschild black hole dressed with a scalar field in four dimensions,  referred to as a Schwarzschild scalarwall. In eleven dimensions, the scalar field arises as moduli of the torus. The resolution occurs at an infinitesimal scale above the horizon, where it transitions to a smooth bubbling spacetime supported by M2-brane flux.

\end{abstract}


\maketitle

The Schwarzschild black hole is the quintessential solution to Einstein equations that underlies some of the most important theoretical and observational results in gravity over the past century.  Two of its most characteristic predictions are the eventual breakdown of general relativity due to the singularity at its center, and the event horizon that hides the singularity from external observers.

While the Schwarzschild black hole is a remarkably simple solution of general relativity,  it is a complex quantum mechanical state that corresponds to the thermodynamic limit of an ensemble of quantum gravity microstates.  As such, it sheds light on many mysteries of quantum gravity and constitutes a theoretical laboratory for exploring the nature of gravity.  This quantum perspective raises two essential questions: first,  what are the degrees of freedom that resolve the singularity? And second,  at what scale do they appear? 

The resolution of the singularity will provide explicit microstate descriptions of black holes.  Although a generic microstate should be quantum mechanical,  some classes could be sufficiently coherent to admit classical descriptions as smooth horizonless solitons of gravity.  Such objects would lead to explicit geometric resolutions of black holes and the emergence of a new gravitational phase of matter.  However,  due to the no-soliton theorem \cite{Serini,Einstein,Einstein:1943ixi,Lichnerowicz}, such a resolution cannot exist in four-dimensional general relativity.  This theorem is avoided by allowing new gravitational degrees of freedom which are extra compact dimensions \cite{Gibbons:2013tqa}.  

Regarding the second question,  a  new physics from quantum gravity may emerge at the scale of the horizon \cite{Mathur:2009hf}.  This can be seen both from a bottom-up analysis as in the firewall paradox \cite{Mathur:2009hf,Almheiri:2012rt} or from a top-down perspective in string theory \cite{Mathur:2005zp,Mathur:2009hf}.  This rather dramatic phenomenon emerges naturally in the microstate geometry program \cite{Bena:2007kg,*Bena:2022rna,*Bena:2022ldq,*Warner:2019jll} where string theory degrees of freedom are used to classically resolve supersymmetric black hole horizons into smooth horizonless geometries.  

The fundamental question we address in this letter is whether such a geometric resolution can also exist for the Schwarzschild black hole.  This is an outstanding problem for more than 20 years in string theory.  Classical resolutions of the Schwarzschild black hole could also have a significant impact on black hole astrophysics in the coming decades.   Such solutions provide a new class of ultra-compact objects that could be detected by the next generation of gravitational wave detectors and imaging telescopes \cite{Barack:2018yly,*Mayerson:2020tpn,*LISA:2022kgy}.

The proposed resolution of the Schwarzschild horizon is based on recent work \cite{Bah:2020ogh,*Bah:2020pdz,*Bah:2021owp,Bah:2021rki,Heidmann:2021cms,Bah:2022yji,Bah:2022pdn,*Heidmann:2022zyd},  in which we develop a new mechanism to systematically construct static topological solitons in generic theories of gravity.  These solitons use geometric transitions induced by the dynamics of extra compact dimensions and supported by electromagnetic flux.  Explicit constructions have been realized by generalizing the Ernst formalism \cite{Heidmann:2021cms,Bah:2022yji,Bah:2022pdn,*Heidmann:2022zyd}.  See \cite{Bah:2021rki,Bah:2021owp} for pedagogical discussion and \cite{Heidmann:2021cms,Bah:2022pdn,Heidmann:2022zyd} for their descriptions as coherent quantum gravity states in string theory.

\begin{figure}[t]
\centering
\includegraphics[width=0.65\textwidth]{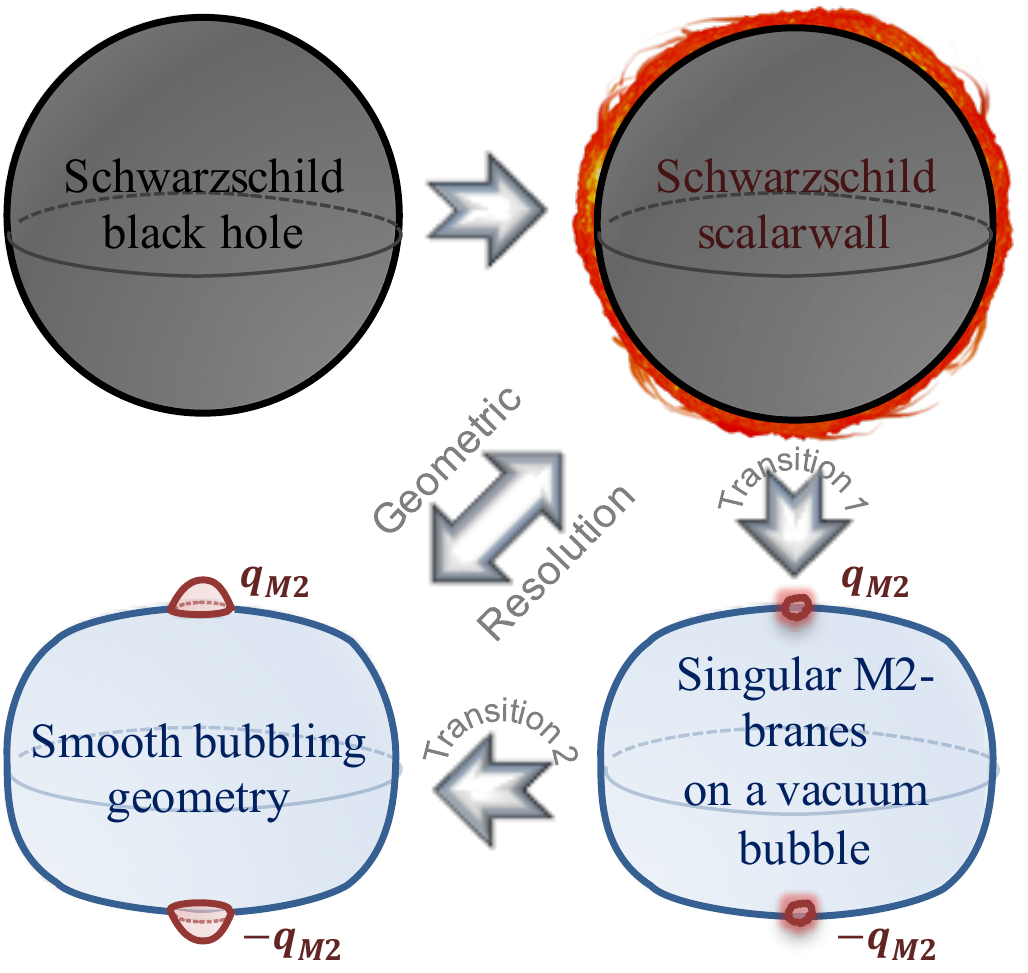}
\caption{Schematic description of the geometric resolution of Schwarzschild horizon.\vspace{-0.4cm}}
\label{fig:Intro}
\end{figure} 

The resolution takes place in  M-theory where the solitonic states are characterized as M2 and anti-M2 ($\overline{\text{M2}}$) brane bound states. It is summarized by the following steps,  schematized in Fig.\ref{fig:Intro}.

{\bf Schwarzschild Scalarwall:} First, we introduce a solution, called \emph{a Schwarzschild scalarwall,} which corresponds to a Schwarzschild black hole dressed with a scalar that has a non-trivial profile at the horizon.  The dressing has an effect of puffing the singularity to the horizon.   The spacetime is well approximated by a Schwarzschild black hole with a singular horizon at $r=2M$ and with even multipole moments that break the spherical symmetry.  

{\bf M-theory Resolution I:} Next, we consider an uplift of the Schwarzschild scalarwalls in M-theory on a seven-torus,  T$^7$,  where they correspond to vacuum solutions.  In this context, the four-dimensional scalar results from deformations of the torus that cause the collapse or blow-up of its inner circles at the horizon.  

We derive a first resolution where the scalarwall horizon is  replaced by a neutral bound state of M2 and $\overline{\text{M2}}$ branes as depicted in Fig.\ref{fig:Intro}.  This is a fully backreacted system where the branes and anti-branes are located at the poles of a vacuum bubble that holds them apart.  The residual dipole structure of this brane bound state provides a microscopic origin for the non-trivial multipole moments of the Schwarzschild scalarwall.  

{\bf M-theory Resolution II:} While the resolution in the previous stage provides a physical microscopic interpretation of Schwarzschild scalarwalls,  the metric is singular at the M2 brane loci.  These regions are further resolved by a geometric transition of the branes into small bubbles supported by flux. We finally obtain smooth topological solitons made of a three-bubble bound state that is indistinguishable from a Schwarzschild scalarwall up to an infinitesimal scale above the horizon at $r=2M$. 

The fact that the resolution occurs for a Schwarzschild scalarwall instead of a Schwarzschild black hole could not be accidental.  Indeed, a direct resolution of Schwarzschild would be subject to the Gregory-Laflamme instability associated with its trivial embedding with extra compact dimensions \cite{Gregory:1993vy}.  The same type of instability does not apply to Schwarzschild scalarwalls due to deformations of the internal compact space.

In section \ref{sec:scalarwallSol},  we define the class of Schwarzschild scalarwalls and derive some of their properties.   In section \ref{sec:M2boundstate} and \ref{sec:3BuSol},  we perform our two steps of brane and geometric transitions resolving the horizon.  In section \ref{sec:GravSig},  we derive some gravitational signatures.

\section{Schwarzschild scalarwalls}
\label{sec:scalarwallSol}

The geometries,  named \emph{Schwarzschild scalarwalls},  are solutions of four-dimensional Einstein-dilaton theory.  They are labeled by their mass $M$ and a parameter $\alpha\geq 0$:
\begin{align}
&\Scale[0.96]{ds_4^2= -f(r)dt^2+H(r,\theta)^{\alpha} \left(\dfrac{dr^2}{f(r)}+r^2 d\theta^2 \right) +r^2\sin^2\theta d\phi^2 ,}\nn \\
&\Scale[0.96]{H(r,\theta) \equiv \left(1+\dfrac{M^2 \sin^2\theta}{r(r-2M)}\right)^{-1}, \quad f(r) \equiv 1-\dfrac{2M}{r}. }\label{eq:FireWS4d}
\end{align} These are supported by a scalar,  $e^{-\Phi} = f(r)^{\sqrt{\alpha}}$.  The Schwarzschild black hole corresponds to the $\alpha=0$ solution where the scalar also decouples.

For a generic $\alpha$,  the timelike Killing vector vanishes at $r=2M$,  defining a horizon.  However,  this locus is singular due to the scalar backreaction which generates a ``scalar wall" on the Schwarzschild black hole.  

\vspace{1.1cm}
\begin{itemize}
\item[•] \underline{Gravitational multipoles:}
\end{itemize}
Gravitational multipole moments are derived using Thorne's approach \cite{Thorne:1980ru,*Bena:2020uup,*Bah:2021jno}.  Unlike Schwarzschild black holes,  Schwarzschild scalarwalls have \emph{non-zero even gravitational multipoles} induced by the scalar backreaction
\begin{equation}
\begin{split}
&\mathcal{M}_{2n} \= c_n M^{2n+1} \,,\\
&c_n = \left(1,\frac{\alpha}{3},\frac{(4-3\alpha)\alpha}{35},\frac{\alpha(8+5\alpha(-4+\alpha))}{231} , \ldots\right),
\end{split}
\end{equation}
where the first element is the mass with $c_0=1$.  

\begin{itemize}
\item[•] \underline{Photon ring and scattering properties:}
\end{itemize}
Massless probes follow geodesic trajectories governed by a potential and two conserved quantities:
\begin{align}
&\dot{r}^2 + r^2 f\,\dot{\theta}^2 - H^{-\alpha} \,V_\text{Sch} =0\,,\quad V_\text{Sch} \equiv 1-\frac{p_\phi^2 f}{r^2\sin^2\theta}\,,\nn\\
&\dot{t} = f^{-1}\,,\qquad \dot{\phi} = \frac{p_\phi^2}{r^2\sin^2\theta}\,,
\end{align}
where $V_\text{Sch}$ corresponds to the Schwarzschild potential.  The effect of the scalarwall is to rescale the Schwarzschild potential by $H^{-\alpha}$.  

The lensing properties as seen from a distant observer are given in Fig.\ref{fig:Imaging} at the end of the letter.  We have used the technique and methodology developed in \cite{Heidmann:2022ehn}.  From these imaging simulations,  we show that Schwarzschild scalarwalls have remarkably similar scattering properties to Schwarzschild.  It has \emph{a shadow},  i.e.  an unstable photon ring,  which is almost spherically symmetric and nearly indistinguishable from the Schwarzschild shadow.  

The shadow can be characterized on the equatorial plane, where its radius is the same as that of Schwarzschild, $R_s = 3M$.  The angular momentum of orbiting photons, $\Omega$, is also the same, while the Lyapunov exponent, $\lambda$,  that determines the instability timescale of the orbits is slightly larger: 
\begin{equation}
p_\phi= \Omega^{-1} =3\sqrt{3}M,\quad \lambda =\left(\frac{2}{\sqrt{3}} \right)^\alpha \lambda_\text{Sch}.
\label{eq:UnstablePhotonSphere}
\end{equation}

Therefore,  Schwarzschild scalarwalls have lensing properties extremely close to those of a Schwarzschild black hole.  Furthermore, since the quasi-normal modes of such objects are intrinsically related to the angular velocity and Lyapunov exponent of photons at the shadow \cite{Cardoso:2008bp},  a similar spectrum of quasi-normal modes is expected.

For the purpose of the resolution, Schwarzschild scalarwalls can be uplifted to vacuum solutions of eleven-dimensional supergravity on T$^7$:
\begin{equation}
\begin{split}
ds^2 &= f(r)^{-\frac{1}{2}\sum_{i=0}^6 D_i}\,ds_4^2 \+ \sum_{i=0}^6 f(r)^{D_i}\, dy_i^2 \,, \label{eq:FireWS}
\end{split}
\end{equation}
where the $D_i$ are seven weights,  and the $\{y_i\}_{i=0,...,6}$ are the coordinates of the T$^7$ with $2\pi R_{y_i}$ periodicity.
The scalarwall field arises from the geometry of the T$^7$ for which its internal circles either blow up ($D_i <0$) or shrink ($D_i>0$) when $f(r)=0$.  In the uplift, we have the identification $\alpha = \frac{1}{2}\sum_{i=0}^6 D_i^2 +\frac{1}{4}\left( \sum_{i=0}^6 D_i\right)^2$.

In this letter,  we restrict to a solution with $\alpha=1$ and
\begin{equation}
D_i=\frac{2}{3}\,,\quad D_j=-\frac{1}{3}\,,\quad i=0,1,2,  \quad j=3,4,5,6.
\label{eq:Weights}
\end{equation}

\section{First geometric transition}
\label{sec:M2boundstate}

The geometric resolution proposed in this letter exploits the eleven-dimensional system in \cite{Heidmann:2021cms} and the mechanism of \cite{Bah:2022yji}.  They allow to replace a spacetime with a horizon,  i.e.  with a shrinking timelike Killing vector, by bubbling geometries where spacelike Killing vectors collapse.

The first transition consists in blowing up a vacuum bubble of size $\ell$ at the place of the horizon.  This is achieved by forcing the $y_0$ circle in T$^7$ to degenerate smoothly.  However, a vacuum bubble does not generate the appropriate redshift to match that of Schwarzschild in four dimensions.  The latter is obtained by adding to the poles of the bubble two singular sources of  M2 and $\overline{\text{M2}}$ branes that are extended along the $(y_1,y_2)$ two-torus. They induce a mass $m$ and have opposite charges so that the solutions remain neutral.  Their strong attraction is supported by the vacuum bubble.  This configuration generates the necessary warping factors to reproduce a Schwarzschild scalarwall up to an infinitesimal scale above the singular horizon (see Fig.\ref{fig:Intro}). 

The M-theory solution is given by
\begin{align}
&\Scale[0.95]{ds^2 =\dfrac{1}{(f^3 Z)^{\frac{1}{6}}} ds_4^2+ Z^\frac{1}{3} \left(\dfrac{f}{U} dy_0^2 + \sum_{i=3}^6 dy_i^2 \right)+\dfrac{U dy_1^2 +dy_2^2}{Z^\frac{2}{3}}, }\nn \\
& F_4= dT \wedge dt \wedge dy_1 \wedge dy_2\,, \label{eq:BuSol}\\
&\Scale[0.96]{ds_4^2= -\sqrt{\dfrac{f}{Z}}dt^2+\sqrt{f Z} \left[H \left(\dfrac{dr^2}{f}+r^2 d\theta^2 \right) +r^2\sin^2\theta d\phi^2\right] ,}\nn
\end{align}
where we have
\begin{align}
f &\equiv 1- \frac{\ell}{r},\quad U=1,\quad  H \equiv 1- \frac{4m^2 \sin^2 \theta}{(2 r-\ell)^2-\ell^2 \cos^2\theta} ,  \nn\\
Z &\equiv 1 + \frac{4 m(2 r-(\ell-2m))}{(2r-\ell)^2-\ell^2 \cos^2 \theta -4 m^2 \sin^2 \theta}\,,\\
T &\equiv \frac{4 m\sqrt{\ell^2-4m^2}\,\cos \theta}{(2r-(\ell-2m))^2-(\ell^2-4m^2) \cos^2 \theta}\,,\nn
\end{align}
with the range $\ell \leq r < \infty$ and $0 \leq 2m \leq \ell$.  

The limit $\ell=2m=2M$ leads to $Z=f^{-1}$ and $T=0$,  which corresponds to the Schwarzschild scalarwall with the weights \eqref{eq:Weights}.  The solution with $\ell \sim 2m $ becomes indistinguishable from the Schwarzschild scalarwall but resolves its singular horizon into a non-trivial brane and  topological microstructure.

For generic values of $\ell$ and $m$,  the spacetime terminates at $r=\ell$.   First,  the Schwarzschild factor $f$ is now produced by the degeneracy of the spacelike $y_0$ direction,  defining a smooth spacetime bubble when \footnote{To derive the regularity condition,  one needs to describe the topology as $r \to \ell$.  We will find a T$^4\times$S$^2$ fibration over an origin of $\IR^2$ of the type $d\rho^2 +C^2 \rho^2 dy_0^2$,  where $\rho \to 0$ and $C$ is a constant that depends on $\ell$ and $m$.  The absence of conical defect requires $C=R_{y_0}^{-1}$.}
\begin{equation}
R_{y_0} = 2\sqrt{\ell^2-4m^2}\,.
\end{equation}
Here,  $R_{y_0}$ is the asymptotic radius of the $y_0$ circle.  The collapsing circle generates the $\sqrt{f}$ in the four-dimensional metric \eqref{eq:BuSol}.  While the electric potential, $T$, due to the M2 branes generate the $\sqrt{Z}$ term.  Taken together, they reproduce the necessary redshift.  

One can check that $Z$ diverges at $r=0$ and $\theta=0,\pi$,  defining the loci of $N$ M2 and $N$ $\overline{\text{M2}}$ branes, thereby leading to \emph{a neutral bound state of M2 and $\overline{\text{M2}}$ branes on a vacuum bubble.} The number of M2 branes at each pole of the bubble is \footnote{It is generically given by $N = (2\pi \ell_p)^{-6} \int \star F_4$.  }
\begin{equation}
N \= \frac{2 m R_{y_0}}{\ell_p^2}\,\sqrt{\frac{\ell+2m}{\ell-2m}}\,\cV_4\,,
\end{equation}
where $\ell_p$ is the Planck length in eleven dimensions and $\cV_4$ is the volume of T$^4$ in Planck unit,  $\cV_4 =\prod_{i=3}^6 R_{y_i}/\ell_p^4$.  

From a four-dimensional perspective upon compactification on the T$^7$,  the bound state corresponds to a singular massive electric dipole,  for which the singularity at $r=\ell$ is resolved through a higher-dimensional spacetime bubble except at its poles.  More precisely,  it reduces to the four-dimensional metric \eqref{eq:BuSol},  a one-form gauge field $T dt$,  and three scalars \cite{Heidmann:2021cms}.  The mass and the electric dipole are given by
\begin{equation}
M = \frac{\ell+2m}{4} \,,\qquad J \= m \sqrt{\ell^2-4m^2}\,.
\end{equation}
We invert the parametrization in terms of $M$ and $R_{y_0}$:
\begin{align}
\ell &= 2M (1+\epsilon_0) \,,\quad m = M(1-\epsilon_0) \,,\quad \epsilon_0\equiv\frac{R_{y_0}^2}{64M^2},\nn \\
 J &= \frac{MR_{y_0}}{2}(1-\epsilon_0)\,,\quad N = \frac{16M^2\cV_4}{\ell_p^2} (1-\epsilon_0)\,. \label{eq:BSquantities}
\end{align}
A macroscopic solution with $M \gg R_{y_0}$ necessarily implies $\ell \sim 2m$,  which is precisely the Schwarzschild scalarwall limit. Thus,  a bound state of large number of M2 and $\overline{\text{M2}}$ branes on a vacuum bubble is very well approximated by the Schwarzschild scalarwall with \eqref{eq:Weights}.  

The scale of resolution is determined by the scale for which $Z^{-1}$ and $f$ start to differ from $1-2M/r$.  This occurs only when $r-2M$ gets smaller than $R_{y_0}$,  so an arbitrarily small scale above the horizon.

One of the remarkable properties is that the geometry,  while being supported by a large electric flux of the order $M^2$, is well approximated by a vacuum solution at a distance of order $R_{y_0}$ above the horizon.   This is made possible by a simple phenomenon: the two charged sources are separated by a large geometric distance $\ell\sim 2M$,  but their physical distance is almost zero due to the strong warping of the spacetime in between them.  Thus, the macroscopic quantities associated with the flux, such as the electric dipole,  are infinitesimal.

\section{Second geometric transition}
\label{sec:3BuSol}

We complete the resolution by resolving the M2 brane singularities.   We blow up two bubbles of size $2\sigma$ at the place of the singular M2-brane sources and support them with similar M2-brane flux (see Fig.\ref{fig:Intro}).  These are generated by forcing  the degeneracy of one of the T$^2$ circles,  i.e. $y_1$ or $y_2$.  The two bubbles are separated by the same vacuum bubble as before but now of size $\ell-2\sigma$.  The addition of the flux will induce mass and charge parameters,  $m$ and $q$, which fix $\sigma$ as
\begin{equation}
\sigma \= \sqrt{m^2-q^2 \,\frac{\ell-2m}{\ell+2m}}\,.
\end{equation}
There is an extremal bound given as $q \leq m \sqrt{\frac{\ell+2m}{\ell-2m}}$.  When saturated,  the bubbles are point-like and degenerate to the M2-brane sources of the previous section.  In a regime where $q$ is near extremal,  the solution is indistinguishable from the previous bound state but resolves the singular M2 branes at a small scale of order $\sigma$ away from the singularities.

The solution in eleven dimensions is given in \eqref{eq:BuSol} with details provided in Appendix \ref{app:3BUSol}.  The spacetime smoothly caps off at $r= \ell+2\sigma$ corresponding to a bubbling region.
We have 
\begin{equation}
f = 1- \frac{\ell+2\sigma}{r}\,,\qquad U = \frac{r_1 r_3}{(r_1+2\sigma)(r_3+2\sigma)}\,,
\end{equation}
where $r_1$ and $r_3$ are the radial distances from the two outer bubbles given in \eqref{eq:LocSphCoor}.  Moreover,  $Z$, $T$ and $H$ have been modified from the previous section such that $Z$ is finite and non-vanishing in the whole spacetime.  They are expressed in \eqref{eq:BuMainFunc}.  

Upon compactification on the T$^7$,  the solution is given by the four-dimensional metric \eqref{eq:BuSol},  a one-form gauge field, and five scalars \cite{Heidmann:2021cms}.  From a four-dimensional perspective,  the solution is singular at $r=\ell+2\sigma$ with a mass and an electric dipole:
\begin{equation}
M \= \frac{\ell+2m}{4}\,,\qquad J= q(\ell-2m)\,.
\end{equation}
The singularity is fully resolved in eleven dimensions,  so that we have either $U=0$ or $f/U=0$ at $r=\ell+2\sigma$ forcing the $y_1$ or $y_0$ circle to degenerate respectively.  The regularity of the local metric requires
\begin{equation}
\begin{split}
R_{y_0}&=\frac{2\sqrt{(\ell^2-4m^2)(\ell^2-4\sigma^2)}}{\ell} \,,\\
k R_{y_1}&=2\sqrt{\frac{(\ell+2m)(m+\sigma)(\ell+2\sigma)}{\ell}}\,, \label{eq:Reg3BU}
\end{split}
\end{equation}
where we have introduced a conical defect $k\in \mathbb{N}$ at the two outer bubbles.  The units of flux that support these two and give rise to the opposite charges are:
\begin{equation}
N = \frac{2 q R_{y_0} \cV_4}{\ell_p^2}\,.
\end{equation}
The internal parameters can be expressed in terms of the asymptotic quantities and the defect by inverting \eqref{eq:Reg3BU}:
\begin{align}
&\medmath{\ell = 2M \left(1+\frac{R_{y_0}^2}{\bar{R}_y (16M-\bar{R}_y)} \right), \quad m = M \left(1-\frac{R_{y_0}^2}{\bar{R}_y (16M-\bar{R}_y)} \right),} \nn \\
&\medmath{ q= \frac{kR_{y_1}\sqrt{(16M-\bar{R}_y)^2-R_{y_0}^2}}{4R_{y_0}},\quad \bar{R}_y^2 \equiv 4k^2 R_{y_1}^2 +R_{y_0}.}
\end{align}
The bound for $q$ restricts the mass to the following range:
\begin{equation}
R_{y_0} + \bar{R}_y \leq 16M \leq 2\bar{R}_y\,.
\label{eq:MassBound}
\end{equation}
The solution describes a macroscopic geometry  with $M \gg R_{y_0}, R_{y_1}$ and with small $\sigma$ if
\begin{equation}
\epsilon_0 \ll 1\,,\quad \epsilon_1 \equiv \frac{k R_{y_1}}{4 M} -1 \ll 1 \quad \Rightarrow\quad k \sim \frac{4M}{R_{y_1}} \gg 1,\label{eq:Regime} \nn
\end{equation}
where $\epsilon_0$ is given in \eqref{eq:BSquantities}.  In this regime,  the singular M2-brane loci of the previous bound states have been replaced by two small charged bubbles with conical defects.  

A conical defect is necessary to scale the mass independent from the radii of the extra dimensions due to the bound \eqref{eq:MassBound}.   Even though a conical defect is strictly speaking a singularity, string theory constructions with orbifolds are generally considered smooth because they can be resolved by known geometric transitions.  First, as shown in \cite{Bah:2020ogh,Bah:2020pdz}, a defect of order $k$ at a bubble can be resolved classically by splitting its poles into $k$ smooth Gibbons-Hawking centers.  Secondly, they can also be resolved by dividing the bubble into $k^4$ microscopic bubbles which will form a ``bubble bag end’' as in \cite{Bah:2021rki}. 

Allowing $k$ in the three-bubble geometry here demonstrates a complete macroscopic resolution, while highlighting important physics, without the need to construct the more complex solution with a large number of bubbles.  It also demonstrates that resolution of macroscopic objects requires many degrees of freedom. 

To conclude,  we have constructed a smooth bubbling geometry that is indistinguishable from the Schwarzschild scalarwall with \eqref{eq:Weights}.  It provides a geometric transition of its horizon into a chain of three bubbles,  the two outer ones are small and are supported by $N$ and $-N$ units of M2-brane flux, respectively.  The middle one is uncharged and prevents the outer ones from collapsing (see Fig.\ref{fig:Intro}).

\section{Gravitational signatures}
\label{sec:GravSig}

Both geometric resolutions of the Schwarzschild scalarwall do not significantly change the macroscopic gravitational signatures discussed in section \ref{sec:scalarwallSol}.  However, due to the internal microstructure, the properties are significantly different near $r=2M$. 

All geometries have a similar unstable photon ring at $r=3M$ as derived in \eqref{eq:UnstablePhotonSphere}.  However, lights that enter the photon sphere of the bubbling geometry will reach the smooth end-to-spacetime at $r\sim2M$ where the redshift factor is very large.  They explore the geometry with chaotic trajectories and exit the structure at a later time.

\begin{figure}[t]
\begin{center}
    {\includegraphics[width=0.95\columnwidth]{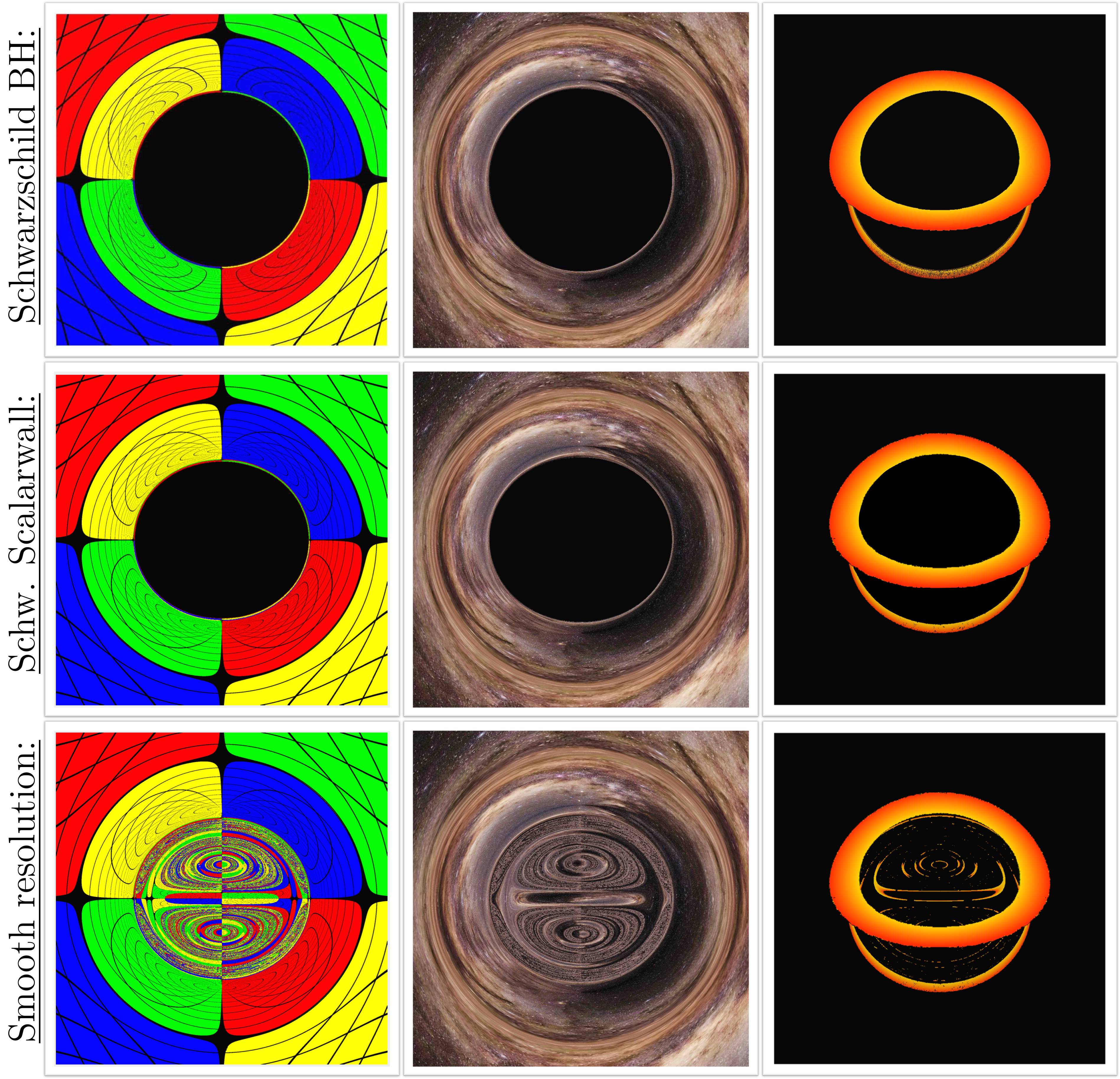}}
\caption{Gravitational lensing simulation of a Schwarzschild black hole, a Schwarzschild scalarwall, and its smooth bubbling resolution of identical mass.   From left to right: the quadricolor screen,  the art-like sky screen,  and the accretion disk picture. \vspace{-0.4cm}}  
\label{fig:Imaging}
\end{center}
\end{figure}

The lensing properties as seen from a distant observer on the equatorial plane are illustrated using the ray tracing code developed in \cite{Heidmann:2022ehn}.  The results and comparison to a Schwarzschild black hole and scalarwall of the same mass are presented in Fig.\ref{fig:Imaging}.  The first column corresponds to the image  when the celestial sphere is covered by a quadricolor grid, which highlights the lensing effect.  The second column provides a more art-like picture when the celestial sphere is covered by a star patch, and the third column gives a simulation of a picture obtained from an accretion disk in the environment near the outer photon sphere.  

First,  Schwarzschild scalarwalls resemble closely a Schwarzschild black hole with a shadow that is slightly squashed at its poles due to the multipolar structure of the solution.  Thus, the difference is undetectable from the imaging of the accretion disk.  

While lights that enter the shadow of the Schwarzschild black hole or scalarwall never escape due to the horizon, lights that enter the resolved geometry exit at a later time.  This provides an important observable that may in the future distinguish ultra-compact objects.  However, incoming photons experience a large delay time and redshift.  As discussed in \cite{Heidmann:2022ehn}, these latter features should lead to a darkening factor that mimics the presence of a horizon.  Therefore, the resolved geometry will be nearly indistinguishable from the Schwarzschild black hole with a blurred and faded light emerging from inside the shadow.

\section{Conclusion}

In this letter, we provided the first example of a geometric transition that resolves the horizon of a Schwarzschild black hole into a smooth bubbling geometry from a top-down perspective in string theory.  The horizon at $r=2M$ has been replaced at an infinitesimal scale by non-trivial topological deformations of spacetime supported by M2 brane flux.  

Ultimately, we would like to explore the stability properties of the solutions derived in this letter.  From \cite{Bah:2021irr,*Stotyn:2011tv}, we know that large flux is required for spacetime bubbles to be meta-stable.  However, unlike \cite{Bah:2021irr,*Stotyn:2011tv}, the present solution, even if supported by large internal flux,  is a bound state of bubbles for which the stability properties may be different.

In addition, we plan to further study the gravitational signatures of Schwarzschild scalarwalls and the geometric resolutions with respect to Schwarzschild.  In a non-exhaustive list, this includes deriving quasi-normal modes, inspiral and ringdown phases in the gravitational signal of binary mergers, and a better model for imaging an accretion disk orbiting the geometries.

\begin{acknowledgments} We are grateful to Iosif Bena, Emanuele Berti and Nicholas P.  Warner for interesting conversations. The work of IB and PH is supported in part by NSF grant PHY-1820784.  The work of IB is also supported in part by the Simons Collaboration on Global Categorical Symmetries.
\end{acknowledgments}

\appendix

\section{The smooth bubbling solution}
\label{app:3BUSol}

The three-bubble solution analyzed in section \ref{sec:3BuSol} is given in terms of spherical coordinates,  $(r,\theta)$ with $r\geq \ell+2\sigma$,  and three sets of local coordinates:
\begin{align}
4 r_1 &\,\equiv\, r_-^{(0)} +r_-^{(1)}-4 \sigma\,, \quad \cos \theta_1 \,\equiv\,  \frac{r_-^{(0)} -r_-^{(1)}}{4\sigma}\,,\nonumber \\
4 r_2 &\,\equiv\, r_-^{(1)} +r_+^{(1)}-2(\ell-2 \sigma)\,, \quad \cos \theta_2 \,\equiv\,  \frac{r_-^{(1)} -r_+^{(1)}}{2(\ell-2 \sigma)}\,,\nonumber \\
4 r_3 &\,\equiv\, r_+^{(1)} +r_+^{(0)}-4 \sigma\,, \quad \cos \theta_3 \,\equiv\,  \frac{r_+^{(1)} -r_+^{(0)}}{4\sigma}\,, \label{eq:LocSphCoor}
\end{align}
where the distance $(r_\pm^{(0)},r_\pm^{(1)})$, are 
\begin{align}
r_\pm^{(0)}& \equiv 2 r - (\ell+2\sigma)(1\pm \cos \theta)\,,  \\
r_\pm^{(1)}  &\equiv\medmath{\sqrt{((2(r-\sigma)-\ell)\cos \theta\pm(2\sigma-\ell))^2+4 r (r-\ell-2\sigma)\sin^2\theta}} .\nonumber
\end{align}
We also define useful constants,
\begin{equation}
\delta \equiv \frac{m^2(\ell+2m)^2+\ell^2 q^2}{(\ell+2m)^2(\ell^2-2m^2)+2\ell^2 q^2}, \quad \gamma \equiv \frac{2mq}{\ell+2m} \,.\nonumber
\end{equation}
The pair $(r_i,\theta_i)$ corresponds to the spherical coordinates centered around the $i^\text{th}$ bubble of the three-bubble bound state.
The warp factors and the gauge potential introduced in \eqref{eq:BuSol} are given by
\begin{widetext}
\begin{align}
& \medmath{ Z=  \frac{(r_1+\sigma + m)(r_3+\sigma +m)+\left(q-\gamma(1+\cos \theta_3)\right)\left(q-\gamma(1-\cos \theta_1)\right)}{\sqrt{\left((r_1+2\sigma)^2+\gamma^2\,\sin^2 \theta_3\,\left(1+\frac{2\sigma}{r_1}\right)\right)\left((r_3+2\sigma)^2+\gamma^2\,\sin^2 \theta_1\,\left(1+\frac{2\sigma}{r_3}\right)\right)}}, } \label{eq:BuMainFunc}\\
& \medmath{ H^4=    \frac{r_1^2 r_2^4 r_3^2 \,(r_2+\ell-2\sigma)^2(r-\ell-2 \sigma)^{-2}}{\left(\left(r_2+\frac{\ell}{2}-\sigma(1-\cos \theta_1) \right)^2-\frac{\ell^2}{4}\right)\,\left(\left(r_2+\frac{\ell}{2}-\sigma(1+\cos \theta_3) \right)^2-\frac{\ell^2}{4}\right)} \left(1+2\delta \,\frac{(q-\gamma)(r_1-r_3)+(\gamma m-\ell(q-\gamma))(\cos \theta_1+\cos \theta_3)}{(q-\gamma)(r_3-r_1)+\gamma m (\cos \theta_1+\cos \theta_3)}\right)^2}  \nn\\ 
&\hspace{0.8cm}\medmath{\times \frac{(r_1(r_1+2\sigma)+\gamma^2\sin^2\theta_3)(r_3(r_3+2\sigma)+\gamma^2\sin^2\theta_1)}{(1+2\delta)^2\,\left(r_1+\sigma(1-\cos \theta_1) \right)^4\left(r_3+\sigma(1+\cos \theta_3) \right)^4} \,, \quad T = \frac{(q-\gamma)(r_1-r_3)-\gamma m (\cos \theta_1 +\cos \theta_3)}{(r_1+\sigma+m)(r_3+\sigma+m)+(q-\gamma(1+\cos \theta_3))(q-\gamma(1-\cos \theta_1))} },\nn 
\end{align}
\end{widetext}

\bibliographystyle{apsrev4-1}
\bibliography{microstates}

\end{document}